\documentstyle[graphicx,amsmath,amsfonts]{article}
\def\be{\begin{equation}}
\def\eea{\end{eqnarray}}
\def\bea{\begin{eqnarray}}
\def\ee{\end{equation}}

\def\r{\rangle}
\def\l{\langle}
\def\nn{\nonumber}

\author{F. Kheirandish$^{1}$ \footnote{fardin$_{-}$kh@phys.ui.ac.ir}, H.
Pahlavani$^{1}$
\\ $^{1}$ {\small Department of Physics, University of Isfahan,}
\\ {\small Hezar Jarib Ave., Isfahan, Iran.}}

\title{Deformed parafermionic algebra from single-band tight-binding
dynamics}

\begin{document}

\maketitle

\begin{abstract}
The quantum dynamics of a driven single-band tight-binding model
with different boundary conditions is considered. The relation
between the Hamiltonian describing the single-band tight-binding
dynamics and the Hamiltonian of a discrete-charge mesoscopic
quantum circuit is elucidated. It is shown that the former
Hamiltonian, with Dirichlet boundary conditions, can be considered
as a realization of the deformed parafermionic polynomial
algebras.
\end{abstract}

\section{Introduction}

 The dynamics of a quantum particle in a periodic potential under
 the effect of an external
field is one of the most fascinating phenomena of quantum physics
[1,2]. Under this condition, the electronic wave function displays
so called Bloch Oscillations (BO) [3,4], the amplitude of which is
proportional to the band width. A suitable orthogonal basis for
investigating structures with periodic potential is using the
localized states. These states are also called the Wannier-Stark
states [5-9] and the nature of this states has a significant
influence on the electronic transport properties of solids. The
electronic BO was observed for the first time in semiconductor
superlattices [10]. In [10,11], the first experimental observation
of BO was reported. Recently, an increasing interest in the dynamics
of BO can be observed [7,11-20].

An explicit time dependence of physical properties can appear
under several conditions. The most obvious case is when the
external field depends on time. In principle, one has to define an
initial state $|\Psi(t=0)\rangle $ and solve the time-dependent
Schr\"{o}dinger equation. This is a partial differential equation
with at least two variables which is separable in some simple
cases [21,22], but rarely solved analytically.

 In particular for the tight-binding model, some analytical expressions have been
derived both for time-dependent and -independent fields [23-25].
Recently in the context of the tight-binding model, a treatment
based on the dynamical Lie algebra is proposed in [26-29]. The
time-dependent Hamiltonian which we will mainly consider here is the
Hamiltonian of a finite system with Dirichlet boundary conditions
defined by

\begin{eqnarray}\label{e1} \hat{H}=G\sum_{j=1}^{N-1}\left(|j\r\l j+1|+|j+1\r\l
j|\right)+F(t)\sum_{j=1}^{N}j|j\r\l j|,
\end{eqnarray}

 where the ket $|j\r$ represents a Wannier
state located on the site $j$. The Hamiltonian (\ref{e1}) with
infinite boundary conditions has been investigated for example in
[15,27]. The Hamiltonian (\ref{e1}) can be treated as the
Hamiltonian of a finite quantum wire, in tight-binding
approximation, under the influence of a time-dependent electric
field. This Hamiltonian appears in the discussion of the quantum
transport through nanoscale conductors [30]. Hamiltonian (\ref{e1})
also appears in models describing the dynamics of cold atoms in
optical lattices in tight-binding approximation [31].

In the present work we elucidate the similarity between the
Hamiltonian (\ref{e1}) with infinite boundary conditions, i.e,
$-\infty\leq j\leq\infty$, and the Hamiltonian describing the
discrete-charge mesoscopic quantum circuits [32,33,34]. In
particular, we show that the Hamiltonian (\ref{e1}) with Dirichlet
boundary conditions can be considered as a realization of the
deformed parafermionic polynomial algebras. Parafermions of order
$p$ (with $p$ being a positive integer) have been introduced in
[35,36,37]. The nature of these particles is such that at most $p$
identical particles can be found in the same state. The usual spin
half fermions correspond to $p=1$. The notion of parafermionic
algebra has been extended by Quesne [38]. The relation between
parafermionic algebras and other algebras has been given in
[39,40,41].

Some extensions of the Hamiltonian (\ref{e1}) are in order. For
example one can add an aperiodic potential to the Hamiltonian , see
for example [42] or add non liner terms as in pendulum model [43].

The layout of the paper is as follows: In section 2, we introduce
the model and discuss in summary the discrete-charge mesoscopic
quantum circuits. In section 3, periodic boundary condition is
discussed. In section 4, a realization of the deformed
parafermionic polynomial algebra is presented. In section 5, the
 time-evolution operator for a finite system
with Dirichlet and periodic boundary conditions is discussed. In
section 6, we present a summary of our results.

\section{The model} The system under study is the derived quantum motion of a charged
particle in a one-dimensional array of single-band quantum wells
in tight-binding approximation and under the action of an
arbitrary time-dependent external field $F(t)$.

The Hamiltonian of such a system can be written as
\begin{equation}\label{e2}
\hat{H}=G\sum_{j}(|j\r\l j+1| + |j+1\r\l j|)+F(t) \sum_{j}j|j\r\l
j|,
\end{equation}
where the ket $|j\rangle$ represents a Wannier state located on site
j. These states fulfill the orthogonality condition $\langle
j|j'\rangle =\delta_{j j'}$. The force $F(t)$ is an arbitrary
time-dependent force and the real parameter $G$ is the
nearest-neighbors coupling strength. In the following we will
investigate the Hamiltonian (\ref{e2}) and its applications in
derived discrete-charge quantum circuits and specially we will show
that this Hamiltonian, with closed boundary conditions, can be
considered as a realization of the generalized parafermionic
polynomial algebras.

\subsection{Infinite chain}
As the first boundary condition let us assume that the Hamiltonian
(\ref{e2}) describes an infinite chain of single-state quantum
wells, i.e, $-\infty <j<+\infty$. In this case the Hamiltonian
(\ref{e2}) can be written as [27]
\begin{equation}\label{e3}
\hat{H}=G(K+K^{\dag})+F(t)\hat{N},
\end{equation}
where
\begin{eqnarray}\label{e4}
\hat{K}&=&\sum_{j=-\infty}^{\infty}|j\rangle\langle j+1|,\nonumber\\
\hat{K}^{\dag}&=&\sum_{j=-\infty}^{\infty}|j+1\rangle\langle j|,\nonumber\\
\hat{N}&=&\sum_{j=-\infty}^{\infty}j|j\rangle\langle j|.
\end{eqnarray}

Operators $K$ and $K^{\dag}$ act as ladder operators on the
Wannier states

\bea\label{e5} \hat{K}|j\rangle
&=& |j-1\rangle,\nonumber\\
\hat{K}^{\dag}|j\rangle &=& |j+1\rangle,
\eea

and the operator $\hat{N}$ acts on the Wannier states as the
position operator

\be\label{e6} \hat{N}|j\rangle=j|j\rangle. \ee

These operators fulfill the following algebra

\bea\label{e7}
&&[\hat{N}, \hat{K}]= -\hat{K},\nonumber\\
&&[\hat{N}, \hat{K}^{\dag}]= \hat{K}^{\dag},\nonumber\\
&&[\hat{K}^{\dag},\hat{K}]=0. \eea

A quantum theory for mesoscopic circuits was proposed by Li and Chen
[32] where charge discreteness was considered explicitly. In this
case the charge operator $\hat{q}$ takes on discrete eigenvalues
when acting on the charge eigenvectors

\be\label{e8} \hat{q}|n\r=nq_{e}|n\r, \ee

where $n\in \cal{Z}$ (set of integers) and $q_{e}$ is the electron
charge. The charge operator $\hat{q}$ can be represented in terms of
the localized states as

\begin{equation}\label{e9} \hat{q}=q_{e}\sum_{j=-\infty}^{\infty}j|j\r\l
j|=q_{e}\hat{N}.
\end{equation}

Discrete-charge mesoscopic quantum circuits are described by the
Hamiltonian

\begin{eqnarray}\label{e10}
&& \hat{H}=-\frac{\hbar^2}{2Lq_{e}^2}(\hat{Q}+\hat{Q}^{\dag}-2)+V(\hat{q}),\nn\\
&& \hat{Q}= e^{\displaystyle\frac{i}{\hbar}q_{e}\hat{p}},\nn\\
&& [\hat{q},\hat{p}]=i\hbar,\nn\\
\end{eqnarray}

where $L$ is the inductance of the circuit. The operators
$\hat{Q}$ and $\hat{Q}^\dagger$ are charge ladder operators which
together with the charge operator $\hat{q}$ fulfill the following
commutation relations

\begin{eqnarray}\label{e11}
&&[\hat{q},\hat{Q}]=-q_{e}\hat{Q},\nonumber\\
&&[\hat{q},\hat{Q}^{\dag}]=q_{e}\hat{Q}^{\dag},\nonumber\\
&&[\hat{Q}^{\dag},\hat{Q}]=0.
\end{eqnarray}

 Therefore the operators $\hat{N}=q_{e}^{-1}\hat{q}$, $\hat{Q}$
 and $\hat{Q}^{\dag}$ satisfy the same algebra (\ref{e7}). When
 $V(\hat{q})=\frac{\hat{q}^2}{2C}$, that is when there is a capacitor
 in the circuit with capacity $C$, we have a $LC$-design and for
 $V(\hat{q})=0$, we have a pure $L$-design.

 The time-dependent Hamiltonian of a L-design circuit in charge representation and
 under the influence of an external potential $\epsilon(t)$ is [32]

 \begin{equation}\label{e12}
 \hat{H}=-\frac{\hbar^{2}}{2Lq_{e}^{2}}(\hat{Q}+\hat{Q}^{\dag})+\epsilon(t)\hat{q},
 \end{equation}

  which is mathematically equivalent to the Hamiltonian (\ref{e3}).

\section{Periodic boundary condition}
We can also take the periodic boundary condition i.e, $|n+N\r=|n\r$.
This kind of boundary condition appears for example in describing a
circular quantum wire in tight-binding approximation. Where, a
time-dependent magnetic field piercing the ring, acts as a driving
force according to the Faraday's law. In this case the Hamiltonian
(\ref{e2}) and the corresponding operators are defined as

\begin{equation}\label{e13} \hat{H}=G\sum_{j=1}^{N}(|j\r\l j+1| + |j+1\r\l
j|)+F(t) \sum_{j=1}^{N}j|j\r\l j|,
\end{equation}

\bea\label{e14}
\hat{K} &=& \sum_{j=1}^{N} |j\r\l j+1|,\nonumber\\
\hat{K^\dag}&=&\sum_{j=1}^{N} |j+1\r\l j|,\nonumber \\
\hat{N}&=& \sum_{j=1}^{N} j |j\r\l j|. \eea

 The commutation relations are now a deformation of the algebra (\ref{e7}) as

\begin{eqnarray}\label{e15}
&&[\hat{N},\hat{K}]=-\hat{K}(1-h(\hat{N})),\nonumber\\
&&[\hat{N},\hat{K}^{\dag}]=(1-h(\hat{N}))\hat{K}^{\dag},\nonumber\\
&&[\hat{K},\hat{K}^{\dag}]=0,\nonumber\\
\end{eqnarray}

where the polynomial operator $h(\hat{N})$ is the deformation
operator defined by

\begin{equation}
h(\hat{N})=\frac{N(-1)^{N-1}}{(N-1)!}\prod_{j=2}^{j=N}(\hat{N}-j)
\equiv N|1\r\l 1|.
\end{equation}

\section{Deformed parafermionic algebra}
In this section we show that the Hamiltonian (\ref{e1}), with
Dirichlet boundary conditions, can be considered as a realization of
the deformed parafermionic polynomial algebra. The Hamiltonian
(\ref{e1}) appears for example in the dynamics of cold atoms in
optical lattices [31,44] and also in the modeling of a finite
quantum wire, in tight-binding approximation and under an external
time-dependent electric field [30]. The Hamiltonian can be written
as

\begin{eqnarray}\label{e17} &&\hat{H}=G\sum_{j=1}^{N-1}(|j\r\l j+1| + |j+1\r\l
j|)+F(t) \sum_{j=1}^{N}j|j\r\l j|,\nonumber\\
&&\hat{K}=\sum_{j=1}^{N-1}|j\r\l j+1|,\nonumber\\
&&\hat{K}^{\dag}=\sum_{j=1}^{N-1}|j+1\r\l j|,\nonumber\\
&&\hat{N}=\sum_{j=1}^{N}j|j\r\l j|.
\end{eqnarray}

The Hamiltonian in terms of the generators can be rewritten as

\begin{equation}\label{e18}
\hat{H}(t)=G(\hat{K}+\hat{K^\dagger})+F(t)\hat{N},
\end{equation}

 the new commutation relations are now as follows

\begin{eqnarray}\label{e19}
&&[\hat{N},\hat{K}]=-\hat{K},\nonumber\\
&&[\hat{N},\hat{K^\dag}]=\hat{K}^{\dag},\nonumber\\
&&[\hat{K}^{\dag},\hat{K}]=|N\r\l N|-|1\r\l 1|.
\end{eqnarray}

From (\ref{e17}), it is clear that the operators $\hat{K}$ and
$\hat{K}^{\dag}$ are nilpotent of order $N$, that is,

\begin{equation}\label{e20}
\hat{K}^{N}=\hat{K}^{\dag N}=0.
\end{equation}

The last commutation relation in (\ref{e19}) can be expressed as a
polynomial in terms of the operator $\hat{N}$. This polynomial is
not necessarily unique and we can choose a polynomial with the
smallest degree. For this purpose let the unknown polynomial be
$f(\hat{N})$ then

\begin{equation}\label{e21}
 f(\hat{N})=|N\r\l N|-|1\r\l 1|.
\end{equation}

 The polynomial $f(\hat{N})$ can be determined from (\ref{e21}) using the following
 conditions

\begin{eqnarray}\label{e22}
f(\hat{N})|n\r &=& 0,\hspace {2 cm}\mbox{for}\hspace{1 cm}2\leq n\leq N-1,\nonumber\\
f(\hat{N})|N\r &=& |N\r,\nonumber\\
f(\hat{N})|1\r &=& -|1\r,\nonumber\\
\end{eqnarray}

after some simple algebra we find

\begin{eqnarray}\label{e23}
f(\hat{N})&=&\frac{1}{(N-2)!
}\prod_{j=2}^{N-1}(\hat{N}-j),\hspace{3 cm}\mbox{for odd $N\geq 3$},\nonumber\\
f(\hat{N})&=&\frac{1}{(N-1)!
}(2\hat{N}-(N+1))\prod_{j=2}^{N-1}(\hat{N}-j),\hspace{0.5cm}\mbox{for
even $N>2$},\nonumber\\
f(\hat{N})&=&(2\hat{N}-3)\hspace{5cm}\mbox{for $N=2$}.\nonumber\\
\end{eqnarray}

Let us define the polynomial operator $g(\hat{N})$ as

\begin{equation}\label{e25}
f({\hat{N}})=g(\hat{N}+1)-g(\hat{N}).
\end{equation}

From (\ref{e23}) one can find $g(\hat{N})$ up to a constant as
follows

\begin{eqnarray}\label{e26}
&&g(\hat{N})=
\frac{1}{(N-1)!}\prod_{j=2}^{N}(\hat{N}-j),\hspace{2.2cm}
\mbox{ for odd $N\geq 3$},\nonumber\\
&&g(\hat{N})=
(\frac{2}{N!}\hat{N}-\frac{N+2}{N!})\prod_{j=2}^{N}(\hat{N}-j),\hspace{1
cm}\mbox{ for even $N\geq 2$}.\nonumber\\
\end{eqnarray}

Putting everything together, we arrive at the following polynomial
algebra

\begin{eqnarray}\label{e27}
&&[\hat{N},\hat{K}^{\dag}]=\hat{K}^{\dag},\nonumber\\
&&[\hat{N},\hat{K}]=-\hat{K},\nonumber\\
&&[\hat{K}^{\dag},\hat{K}]=f(\hat{N})=g(\hat{N}+1)-g(\hat{N}),\nonumber\\
&&[\hat{K}^{\dag}\hat{K},\hat{K}\hat{K}^{\dag}]=0,\nonumber\\
&&\hat{K}^{N}=\hat{K}^{\dag N}=0,\nonumber\\
&&\hat{K}^{\dag}\hat{K}=1-g(\hat{N})=:\phi(\hat{N}),\nonumber\\
&&\hat{K}\hat{K}^{\dag}=1-g(\hat{N}+1)=:\phi(\hat{N}+1),\nonumber\\
\end{eqnarray}

where

\begin{equation}\label{e28}
\phi(1)=\phi(N+1)=0,
\end{equation}

as is clear from the definition of the operators $\hat{K}$ and
$\hat{K^{\dag}}$. Now we show that the polynomial algebra
(\ref{e27}) is in fact a deformation of the usual parafermionic
polynomial algebra ${\mathcal{L}}=\{\hat{B},\hat{B^\dag},\hat{M}\}$
[35] defined by

\begin{eqnarray}\label{e29}
&&[\hat{M},\hat{B}]=-\hat{B},\nonumber\\
&&[\hat{M},\hat{B}^{\dag}]=\hat{B}^{\dag},\nonumber\\
&&\hat{B}^{p+1}=\hat{B}^{\dag p+1}=0,\nn\\
&&\hat{B}^{\dag}\hat{B}=\hat{M}(p+1-\hat{M})=[\hat{M}],\nn\\
&&\hat{B}\hat{B}^{\dag}=(\hat{M}+1)(p-\hat{M})=[\hat{M}+1],\nn\\
&&\hat{M}=\frac{1}{2}([\hat{B}^{\dag},\hat{B}]+p),\nn\\
\end{eqnarray}

where $p$ is a nonzero positive integer. For this purpose let us
define the operators $\hat{B}$, $\hat{B^{\dag}}$ and $\hat{M}$ as
\begin{eqnarray}\label{deform}
\hat{B}&=&\sqrt{\hat{N}(N-\hat{N})}\hat{K},\nonumber\\
\hat{B^{\dag}}&=&\hat{K^{\dag}}\sqrt{\hat{N}(N-\hat{N})},\nonumber\\
\hat{M}&=&\hat{N}-1,\nonumber\\
\end{eqnarray}

then it can be easily shown that the operators $\hat{B}$,
$\hat{B^{\dag}}$ and $\hat{M}$, fulfill the same algebra
(\ref{e29}). It is interesting that for the special cases $N=2,3$
the algebra (\ref{e27}) coincides with (\ref{e29}) and for $N\geq
4$, it is a deformmation of the algebra (\ref{e29}) with the
deformation factor $\sqrt{\hat{N}(N-\hat{N})}$ [38]. The deformation
factor is not invertible since $\sqrt{\hat{N}(N-\hat{N})}|N\rangle
=0$ but we can make it invertible by introducing an infinitesimal
parameter $\epsilon$ and defining a new operator
$\hat{D}_{\epsilon}:=\sqrt{\hat{N}(N+\epsilon-\hat{N})}=
\sqrt{(\hat{M}+1)(N-1+\epsilon-\hat{M})}$ which tends to the
deformation factor when $\epsilon\rightarrow 0$. Now the Hamiltonian
can be written in terms of the generators of the parafermionic
polynomial algebra (\ref{e29}) as
\begin{equation}
\hat{H}=G(\hat{D}^{-1}_{\epsilon}(\hat{M})\hat{B}+\hat{B}^{\dag}
\hat{D}^{-1}_{\epsilon}(\hat{M}))+F(t)(\hat{M}+1).
\end{equation}
The propagator for the non deformed Hamiltonian
\begin{equation}
\hat{H}^{*}=G(\hat{B}+\hat{B}^{\dag})+F(t)(\hat{M}+1),
\end{equation}
can be obtained exactly from the dynamical lie algebra method [45].
\subsection{The $su(2)$ representation of the special cases $N=2,3$}
Setting $N=2$ in equations (\ref{e27}), ( \ref{e23}) we arrive at

\begin{eqnarray}\label{e32}
&&[\hat{N},\hat{K}]=-\hat{K},\nonumber\\
&&[\hat{N},\hat{K}^{\dag}]=\hat{K}^{\dag},\nonumber\\
&&[\hat{K}^{\dag},\hat{K}]=2\hat{N}-3,\nonumber\\
&&\hat{K}^{2}=\hat{K}^{\dag 2}=0,\nonumber\\
&&\hat{N}=1+\hat{K^\dag}\hat{K}.\nonumber\\
\end{eqnarray}

From $\{\hat{K},\hat{K}^{\dag}\}=I$ and $\hat{K}^{2}=\hat{K}^{\dag
2}=0$, it is clear that the set $\{\hat{K},\hat{K}^{\dag},I\}$ is
a realization of the usual spin-half fermionic algebra. By
defining

\begin{eqnarray}\label{e33}
&&\hat{J_{0}}=\hat{N}+\frac{3}{2},\nonumber\\
&&\hat{J_{+}}=\hat{K^\dag},\nonumber\\
&&\hat{J_{-}}=\hat{K},\nonumber\\
\end{eqnarray}

we find

\begin{eqnarray}\label{e34}
&&[\hat{J_{0}},\hat{J_{\pm}}]=\pm\hat{J_{\pm}},\hspace {2.5 cm}
[\hat{J_{+}},\hat{J_{-}}]=2\hat{J_{0}},
\end{eqnarray}

which is the usual $su(2)$ algebra. In terms of $\hat{J_{\pm}}$
and $\hat{J_{0}}$, the Hamiltonian (\ref{e18}) can be rewritten as

\begin{equation}\label{35}
\hat{H}(t)=G(\hat{J_{+}}+\hat{J_{-}})+F(t)(\hat{J_{0}}-\frac{3}{2}).
\end{equation}

For this Hamiltonian the Casimir operator
$\hat{J}^{2}=\hat{J_{0}}^{2}+\frac{1}{2}(
\hat{J_{+}}\hat{J_{-}}+\hat{J_{-}}\hat{J_{+}})$ is a constant of
motion.

Setting $N=3$ in equations (\ref{e27}), ( \ref{e23}) we find

\begin{eqnarray}\label{e36}
&&[\hat{N},\hat{K}]=-\hat{K},\nonumber\\
&&[\hat{N},\hat{K}^{\dag}]=\hat{K^\dag},\nonumber\\
&&[\hat{K}^{\dag},\hat{K}]=\hat{N}-2,\nonumber\\
&&\hat{K}^{3}=\hat{K}^{\dag 3}=0,\nonumber\\
&&\hat{N}=1+\hat{K}^{\dag}\hat{K}+\hat{K}^{\dag 2}\hat{K}^{2},\nonumber\\
\end{eqnarray}

again we can define the $su(2)$ generators as

\begin{eqnarray}\label{e37}
&&\hat{J_{+}}=\sqrt{2}\hat{K^\dag},\nonumber\\
&&\hat{J_{-}}=\sqrt{2}\hat{K},\nonumber\\
&&\hat{J_{0}}=\hat{N}-2.
\end{eqnarray}

The Hamiltonian in terms of this generators becomes

\begin{equation}\label{e38}
\hat{H}(t)=\frac{G}{\sqrt{2}}(\hat{J_{+}}+\hat{J_{-}})+F(t)(\hat{J_{0}}-2).
\end{equation}
For $N=4,5$, $f(\hat{N})$ is a cubic polynomial operator (like Higgs
algebra) and in general, for $N=n,n+1$, $f(\hat{N})$ is a polynomial
operator of degree $n-1$.
\section{Time-evolution operator}
For $N=2,3$ Hamiltonian has $su(2)$ symmetry and the time-evolution
operator can be obtained exactly. In this case the time-evolution
operator $U(t)$ can be written as
\begin{equation}
U(t)=e^{-\frac{\imath}{\hbar}\mu(t)}e^{-\frac{\imath}{\hbar}f(t)\hat{j}_0}
e^{-\frac{\imath}{\hbar}g(t)\hat{j}_{+}}e^{-\frac{\imath}{\hbar}k(t)\hat{j}_{-}}.
\end{equation}
Inserting the above expression in the following evolution equation
\begin{equation}
\imath\hbar\frac{d}{d t}U(t)=\hat{H}U(t),
\end{equation}
 we can find $U(t)$ straightforwardly [45].

 For $N\geq 4$, the previous method is not successful and we find a perturbative
 expansion for $\hat{U}(t)$ in this case. For this purpose let us write the
 time-evolution operator as
\begin{equation}\label{38.1}
U(t)=e^{-\frac{\imath}{\hbar}\varphi(t)\hat{N}}V(t),
\end{equation}
where $\varphi(t)=\int_{0}^{t}F(t')dt'$. Substituting
Eq.(\ref{38.1}) into the evolution equation
\begin{equation}\label{e39}
\imath\hbar\frac{d}{dt}U(t)=\hat{H}U(t),
\end{equation}

we find the following equation for $V(t)$
\begin{equation}\label{eqV}
\imath\hbar\frac{d}{dt}V(t)=G(e^{-\frac{\imath}{\hbar}\varphi(t)}\hat{K}+
e^{\frac{\imath}{\hbar}\varphi(t)}\hat{K}^{\dag})V(t),
\end{equation}
where the following relations have been used
\begin{eqnarray}\label{baker}
&&e^{\frac{\imath}{\hbar}\varphi(t)\hat{N}}\hat{K}
e^{-\frac{\imath}{\hbar}\varphi(t)\hat{N}}=e^{-\frac{\imath}{\hbar}\varphi(t)}\hat{K},
\nonumber\\
&&e^{\frac{\imath}{\hbar}\varphi(t)\hat{N}}\hat{K}^{\dag}
e^{-\frac{\imath}{\hbar}\varphi(t)\hat{N}}=e^{\frac{\imath}{\hbar}\varphi(t)}
\hat{K}^{\dag}.\nonumber\\
\end{eqnarray}
The relations (\ref{baker}) can be obtained directly from the
Baker-Hausdorf formula. The Eq.(\ref{eqV}) can be written in terms
of the state vector $|\psi(t)\r=V(t)|\psi(0)\r$ as
\begin{equation}\label{state}
\imath\hbar\frac{d}{dt}|\psi(t)\r=G(e^{-\frac{\imath}{\hbar}\varphi(t)}\hat{K}+
e^{\frac{\imath}{\hbar}\varphi(t)}\hat{K}^{\dag})|\psi(t)\r.
\end{equation}
 The new Hamiltonian
\begin{equation}
G(e^{-\frac{\imath}{\hbar}\varphi(t)}\hat{K}+
e^{\frac{\imath}{\hbar}\varphi(t)}\hat{K}^{\dag}),
\end{equation}
has the following eigenvalues and eigenvectors respectively
\begin{eqnarray}\label{e48}
&&\lambda_m=2G\cos\omega_{m},\hspace{2cm} \mbox{$m=1,2\cdots
N$},\nonumber\\
&&|\lambda_m,t\r=d_{m}\sum _{ n=1} ^{ N}e^{\imath n\varphi(t)} \sin
(n\omega_{m})|n\r,
\end{eqnarray}
where $\omega_{m}=\frac{m\pi}{N+1}$ and $d_{m}=\frac{1}{\sqrt{\sum
_{ n=1} ^{ N} \sin^{2} n(\omega_{m})}}$, is a normalization
coefficient. Therefore, the eigenvalues are independent from the
driving force $F(t)$ but the eigenvectors are time dependent and
make an orthonormal complete basis at any time. Now let us expand
the state vector as
\begin{equation}\label{e50}
|\psi(t)\r=\sum _{ \ell=1} ^{ N}C_{\ell}(t)|\lambda_\ell,t\r.
\end{equation}
Inserting (\ref{e50}) into (\ref{state}) we find
\begin{equation}\label{e51}
i\hbar\dot{C_{\ell}}(t)=\hbar\dot{\varphi}\sum _{ \ell'=1} ^{
N}A_{\ell\ell{'}}C_{\ell{'}}(t)+\lambda_\ell(t)C_{\ell}(t).
\end{equation}
where
\begin{equation}\label{e52}
A_{\ell\ell{'}}= \l \lambda_\ell(t)|\frac{d}{dt}
|\lambda_\ell{'}(t)\r=\frac{\sum _{ n=1} ^{ N}n \sin
(n\omega_{\ell}) \sin (n\omega_{\ell{'}})}{\displaystyle\sqrt{\sum
_{ n=1} ^{ N} \sin^{2} (n\omega_{\ell})\sum _{ n=1} ^{ N}\sin^{2}
(n\omega_{\ell{'}})}}
\end{equation}
are the elements of a symmetric matrix $\hat{A}$. The Eq.(\ref{e51})
can be written in matrix form as
\begin{equation}\label{e53}
i\hbar\dot{C}(t)=\hbar\dot{\varphi}\hat{A}C(t)+\hat{\lambda}C(t).
\end{equation}
where

\be\label{e54} C=\left[\begin{array}{cc}
C_1(t)\\C_2(t)\\C_3(t)\\\vdots\\C_N(t)
\end{array}\right]\hspace{0.5cm}\hat{\lambda}=2G\left[\begin{array}{ccccc}
\cos(\omega_{1}) & 0 & 0 & \cdots &  0 \\
0 & \cos(\omega_{2}) & 0 & \cdots &  0\\
0 & 0 & \cos(\omega_{3}) & \cdots &  0\\
\vdots & \vdots & \vdots &\ddots &  \vdots\\
0 & 0 & 0 &  \cdots & \cos(\omega_{N}) \\
\end{array}\right]. \ee
By rescaling the parameters we can assume $2G<1$ and since
$\cos(\omega_{m})<1,\hspace{0.5cm}(1\leq m\leq N)$, so the matrix
$\hat{\lambda}$ has the property
$\lim_{k\rightarrow\infty}\hat{\lambda}^{k}\rightarrow 0$. Now the
equation (\ref{e53}) is similar to the Schr\"{o}dinger equation with
the perturbation $\hat{\lambda}$. Therefore the solution of
(\ref{e53}) can be written as a series expansion in terms of the
matrix $\hat{\lambda}$ as
\begin{equation}\label{e55}
\hat{C}(t)=C^{0}(t)+\hat{\lambda} C^{1}(t)+\hat{\lambda}^{2}
C^{2}(t)+\hat{\lambda}^{3} C^{3}(t)+\cdots,
\end{equation}
substituting (\ref{e55}) into (\ref{e53}) and equating the terms of
the same order in $\hat{\lambda}$ we find
\begin{eqnarray}\label{e56}
\imath\hbar \dot{C}^{0}(t)&=&\hbar\dot{\varphi}\hat{A}C^{0}(0),\nonumber\\
\imath\hbar \dot{C}^{1}(t)&=& {C}^{0}(t)
+\hbar\dot{\varphi}\hat{\lambda}^{-1}\hat{A}\hat{\lambda}{C}^{1}(t),\nonumber\\
&\vdots&\nonumber\\
\imath\hbar \dot{C}^{(n)}(t)&=& {C}^{(n-1)}(t)
+\hbar\dot{\varphi}\hat{\lambda}^{-n}\hat{A}\hat{\lambda}^{n}{C}^{(n)}(t),\nonumber\\
\end{eqnarray}
The first equation leads to
\begin{equation}\label{zero}
C^{(0)}(t)=e^{-\imath\varphi(t)\hat{A}}C^{(0)}(0).
\end{equation}
For $n\geq 1$, let us define
$\hat{\Delta}_n:=\hat{\lambda}^{-n}\hat{A}\hat{\lambda}^{n}$, then
\begin{equation}\label{e57}
(\frac{d}{dt}+i\dot{\varphi}\hat{\Delta}{_{n}})C^{n}(t)=-\frac{i}{\hbar}C^{(n-1)}(t).
\end{equation}
The equation(\ref{e57}) can be solved using the Green function
method and treating $C^{(n-1)}(t)$ as the source term for
$C^{(n)}(t)$ as
\begin{equation}\label{e58}
C^{n}(t)=\exp(-i\varphi(t)\hat{\Delta}{_{n}})C^{n}(0)-\frac{i}{\hbar}
\int_{0}^{t}\exp(-i\varphi(t-t')\hat{\Delta}{_{n}})C^{(n-1)}(t')dt'.
\end{equation}
As the boundary condition we can assume $C(0)=C^{0}(0)$ and
$C^{(k)}(0)=0$ for $k\geq 1$. So the equation (\ref{e58}) can be
simplified as
\begin{equation}\label{e59}
C^{n}(t)=-\frac{i}{\hbar}
\int_{0}^{t}\exp(-i\varphi(t-t')\hat{\Delta}{_{n}})C^{(n-1)}(t')dt',\hspace{1cm}
(n\geq 1).
\end{equation}
Now having (\ref{zero}), we can find $C^{(n)}(t)$ up to any order
recursively.
\section{Conclusions}
The similarity between the Hamiltonian of a single-band
tight-binding model and the discrete-charge mesoscopic quantum
circuits is explicitly shown. It is shown that the former
Hamiltonian, with Dirichlet boundary conditions, can be considered
as a realization of the deformed parafermionic polynomial algebras.


\begin{thebibliography}{99}
\bibitem{[1]} C. Zener, Proc. R. Soc. Lond, A145, {\it Proc. R. Soc. Lond. A}
145, 523 ( 1934)
\bibitem{[2]} F. Bloch, {\it Z. Phys}, 52 555 ( 1928)
\bibitem{[3]} G. H. Wannier, {\it Phys. Rev} 100 ,1227 ( 1955) {\it
Phys. Rev} 101, 1835 ( 1956) {\it Phys. Rev} 117,423 ( 1960)
\bibitem{[4]} D. H. Dunlap and V. M. Kenkre, {\it Phys. Rev. B
}34, 3625 (1986)
\bibitem{[5]} S. Wimberger, R. Mannella, O. Morsch, E. Arimondo, A. R. Kolovsky and
A. Buchleitner {\it Phys. Rev. A} 72, 063610 (2005)
\bibitem{[6]} A. Buchleitner and A. R. Kolovsky, {\it Phys. Rev. Lett} 91,253002 (2003)
\bibitem{[7]} A. R. Kolovsky, {\it Phys. Rev. Lett} 90,213002 (2003)
\bibitem{[8]} M. Gl\"{u}ck, A. R. Kolovsky and H. J. Korsch, {\it Phys. Rep} 366,103 (2002)
\bibitem{[9]} S. Wimberger, P. Schlagheck and R. Mannella  , {\it J. Phys. B} 39,729 (2006)
\bibitem{[10]} K. Leo, P. Hairing, F. Bruggemann, R. Schwedler, and K. Kohler
{\it Solid State Commun.} 84, 943 (1992)
\bibitem{[11]} J. Feldmann, K. Leo, J. Shah, B. A. B. Miller,J. E. Cunningham, T. Meier, G. Von
Plesses, A. Schulze, P. Thomas, and S. S. Schmit-Rink {\it
Phys.Rev. B} 46, 7252 (1992)
\bibitem{[12]} J. B. Krieger and G. J. Iafrate {\it
Phys. Rev. B} 33, 5494 (1985)
\bibitem{[13]} M. Gluck, F. Keck and H. J. Korsch {\it Phys. Rev. A }
66,043418 ( 2002)
\bibitem{[14]} D. Witthaut, M. Werder, S. Mossmam and H. J. Korsch {\it
e-print cond-mat/0403205} (2004)
\bibitem{[15]} T. Hartmann, F. Keck, H. J. Korsch and S. Mossmann, {\it New J.Phys
}6,2 ( 2004)
\bibitem{[16]} G. H. Wannier, {\it Phys. Rev} 117, 432 ( 1960)
\bibitem{[17]} A. R. Kolovsky and H. J. Korsch {\it e-print cond-mat/0403205} ( 2004)
\bibitem{[18]} A. R. Kolovsky, A. V. Ponomarev and H. J. Korsch
{\it e-print quant-ph/0206108}( 2002)
\bibitem{[19]} M. Cristiani, O. Morsch, J. H. Muller, D. Ciampini
and E. Arimondo, {\it Phys. Rev. A} 65, 063612 ( 2002)
\bibitem{[20]} T. Pertsch, T. Zentgraf, U. Peschel, A. Brauer and F. Lederer,{\it
Phys. Rev. Lett} 88, 093901 ( 2002)
\bibitem{[21]} J. R. Ray, {\it Phys. Rev. A} 26,729 (1982)
\bibitem{[22]} A. N. Seleznyova, {\it Phys. Rev. A}51, 950 (1995)
\bibitem{[23]} M. Grifoni and P. Hanggi, {\it Phys. Rep} 304, 229 (1998)
\bibitem{[24]} I. Goychuk, M. Grifoni and P. Hanggi, {\it Phys. Rev. Lett} 81, 649 (1998)
\bibitem{[25]} K.Drese and M.Holthaus, {\it Phys. Rev. Lett} 78,
2932 ( 1996)
\bibitem{[26]} F. Wolf, H. J. Korsch, {\it Phys. Rev. A} 37, 1934 (1988)
\bibitem{[27]} H. J. Korsch, S. Mossmann, {\it Phys. Lett. A} 317, 54-63 (2003)
\bibitem{[29]} H. J. Korsch, {\it Phys. Lett. A} 74, 294 (1979)
\bibitem{[29]8} F. Wolf, {\it J. Phys. A} 20, L421 ( 1987)
\bibitem{[30]} S. Kohler, J. Lehmann and P. H\"{a}nggi, {\it Phy.
Rep.} 406, 379 (2005)
\bibitem{[31]} D. Jaksch et al, {\it Phys. Rev. Lett.} 81, 3108 (1998);
 R. Khomeriki, S. Ruffo and S. Wimberger, {\it e-print
con-mat/0610014} (2006);
\bibitem{[32]} Y. Q. Li and B. Chen {\it Phys. Rev. B} 53, 4027
(1995)
\bibitem{[33]} C. A. Utreras-Diaz and J. C. Flores, {\it e-print
cond-mat/0605141} (2006)
\bibitem{[34]} J. C. Flores and E. Lazo, {\it e-print
cond-mat/9910101} (1999)
\bibitem{[35]} D. Bonatsos, C. Daskaloyannis and K. Kokkotas, {\it e-print
hep-th/9909002 } (1999)
\bibitem{[36]} H. S. Green,{\it Phys. Rev} 90, 270 (1953)
\bibitem{[37]} O. W. Greenberg and A. M. L. Messiah, {\it Phys. Rev. B} 138, 1155 (1965)
\bibitem{[38]} C. Quesne,{\it Phys. Lett. A} 193, 245 (1994)
\bibitem{[39]} D. Bonatsos, C. Daskaloyannis, {\it Phys. Lett. B}
307, 100 (1993)
\bibitem{[40]} N. Debergh  {\it J. Phys. A} 28, 4945 (1995)
\bibitem{[41]} D. Bonatsos, P. Kolokotronis and  C. Daskaloyannis,
{\it Mod Phys. Lett. A} 10, 2197 (1995)
\bibitem{[42]} F. A. B. F. de Moura, M. L. Lyra, F.
Dom\'{i}nguez-Adame, and V. A. Malyshev, {\it Phys. Rev. B} 71,
104303 (2005)
\bibitem{[43]} A. Trombettoni and A. Smerzi, {\it Phys. Rev. Lett.} 86, 2353 (2001);
 A. R. Kolovsky and H. J. Korsch, {\it Int. J. Mod. Phys.}
18, 1235 (2004)

\bibitem{[44]} K. W. Madison, M. C. Fischer, R. B. Diener, Qian Niu, and M. G.
Raizen, {\it Phys. Rev. Lett.} 81, 5093 (1998)
\bibitem{[45]} G. Dattoli, P. Di Lazzaro, {\it Phys. Rev. A} 35, 1582
(1987); C. M. Cheng and P. C. W. Fung, {\it J. Phys. A: Math. Gen.}
21 4115 (1988)

\end{thebibliography}
\end{document}